\documentclass[singlecolumn,pra,superscriptaddress,showpacs,aps,10pt,floatfix]{revtex4-2}
\usepackage{lineno,mathtools,url}
\usepackage{mathbbol} 
\usepackage{physics}
\usepackage{xfrac}
\usepackage{bbold}
\usepackage{gensymb}
\usepackage{multirow}
\usepackage{array}
\usepackage{siunitx}
\usepackage{nicefrac}
\usepackage{amsmath}
\usepackage{amssymb,mathrsfs}
\usepackage{subcaption}

\usepackage{tikz}
\usetikzlibrary{shapes.geometric, arrows.meta, positioning}

\newcommand{\tp}{\mathrm{p}}
\newcommand{\tpr}{\mathrm{pr}}

\providecommand{\vbg}{v_0}         
\providecommand{\nbg}{n_0}         
\providecommand{\deff}{\delta}     
\providecommand{\cs}{c_\mathrm{B}} 
\providecommand{\Lbdg}{\mathcal{L}}

\newcolumntype{C}{>{$}c<{$}}

\begin{document}

\title{Stimulated Hawking effect and quasinormal mode resonance in a polariton simulator of field theory on curved spacetime}

\author{Mattheus Burkhard}
\affiliation{Pitaevskii BEC Center, CNR-INO 
and Dipartimento di Fisica, Universit\`a di Trento, 38123 Trento, Italy}
\affiliation{Université Paris Cité, CNRS, Matériaux et Phénomènes Quantiques, 75013 Paris, France}
\author{Malte Kroj}\affiliation{Laboratoire Kastler Brossel, Sorbonne Universit\'{e}, CNRS, ENS-Universit\'{e} PSL, Coll\`{e}ge de France, Paris 75005, France}
\author{K\'evin Falque}\affiliation{Laboratoire Kastler Brossel, Sorbonne Universit\'{e}, CNRS, ENS-Universit\'{e} PSL, Coll\`{e}ge de France, Paris 75005, France}
\author{Alberto Bramati}\affiliation{Laboratoire Kastler Brossel, Sorbonne Universit\'{e}, CNRS, ENS-Universit\'{e} PSL, Coll\`{e}ge de France, Paris 75005, France}
\author{Iacopo Carusotto}
\affiliation{Pitaevskii BEC Center, CNR-INO 
and Dipartimento di Fisica, Universit\`a di Trento, 38123 Trento, Italy}
\author{Maxime J Jacquet}\email{maxime.jacquet@lkb.upmc.fr}\affiliation{Laboratoire Kastler Brossel, Sorbonne Universit\'{e}, CNRS, ENS-Universit\'{e} PSL, Coll\`{e}ge de France, Paris 75005, France}

\begin{abstract}
    The Hawking effect amplifies fluctuations in the vicinity of horizons, both in black holes and in analogue platforms.
    Here, we consider a polariton simulator and numerically examine the \emph{stimulated} Hawking effect using a coherent probe incident on the horizon from the exterior.
    We implement an experimentally realistic effective spacetime that supports a quasinormal mode (QNM) in the vicinity of the horizon.
    We find that the stimulated Hawking effect manifests as transmission into a negative-energy Bogoliubov channel inside the horizon, consistent with pseudo-unitary Bogoliubov scattering.
    Moreover, transmission across the horizon peaks at the QNM frequency.
    The computed spectral signatures provide a practical guide for future experimental investigations of the Hawking effect and its interplay with QNMs, an open question in quantum field theory in curved spacetime.
\end{abstract}

\maketitle

\section*{Introduction}

In recent years, analogue-gravity systems have provided a compelling platform for studying quantum-field dynamics in curved spacetimes~\cite{barcelo_analogue_2011,jacquet_next_2020,almeida_analogue_2023}.
Such experiments have brought within laboratory reach phenomena traditionally confined to astrophysical scales, most notably the Hawking effect~\cite{philbin_fiber-optical_2008,drori_observation_2019,rousseaux_observation_2008,weinfurtner_measurement_2011,euve_observation_2016,munoz_de_nova_observation_2019}.
Although early theoretical work focused on the scattering of vacuum fluctuations in both conservative and driven–dissipative quantum fluids~\cite{carusotto_numerical_2008,balbinot_nonlocal_2008,Recati_2009,macher_black/white_2009,larre_quantum_2012,Busch_entanglementHR_2014,finazzi_entangled_2014,nova_entanglement_2015,boiron_quantum_2015,finke_observation_2016,robertson_assessing_2017,fabbri_momentum_2018,isoard_departing_2020,isoard_bipartite_2021,ciliberto_bell_2024,Solnyshkov,Gerace,busch_spectrum_2014,jacquet_analogue_2022,Grissins,robertson_four-wave_2019}, experiments have also probed these systems with externally injected fields to \emph{stimulate} the Hawking process and to obtain frequency-resolved insights into the underlying physics~\cite{philbin_fiber-optical_2008,rousseaux_observation_2008,weinfurtner_measurement_2011,choudhary_efficient_2012,drori_observation_2019}.

In this work, we study the stimulated Hawking process at a horizon in a polaritonic quantum fluid of light by injecting a weak monochromatic field.
We numerically model a pump-probe configuration in which a downstream support field maintains an approximately constant density, a regime previously identified as favourable for spontaneous emission.
In contrast to earlier work dominated by vacuum radiation, the probe provides a frequency-resolved read-out of scattering in the asymptotic modes.
We find a crossover from two-port \emph{unitary} scattering (no negative-norm channel available) at $\omega>\omega_\mathrm{max}$ to \emph{pseudo-unitary} scattering with negative-norm participation at $\omega<\omega_\mathrm{max}$, with $\omega_\mathrm{max}$ determined by the transcritical flow~\cite{falque_polariton_2025}.
Here, $\omega_\mathrm{max}$ denotes the upper frequency bound for which mixed-norm scattering channels exist asymptotically; above $\omega_\mathrm{max}$ only positive-norm channels remain, and Hawking-type amplification is unavailable.
Moreover, transmission into the supercritical region exhibits a pronounced resonance at the quasinormal-mode frequency $\omega_{\mathrm{qnm}}>\omega_\mathrm{max}$, i.e. outside the Hawking-emission window~\cite{jacquet_quantum_2023}.
This enables a direct comparison between horizon and horizonless emission---expected to display distinct correlation structures~\cite{jacquet_influence_2020}---and recasts the QNM as an active mediator of stimulated scattering rather than a passive geometric fingerprint.

\begin{figure}[hb!]
    \centering
    \includegraphics[width=\linewidth]{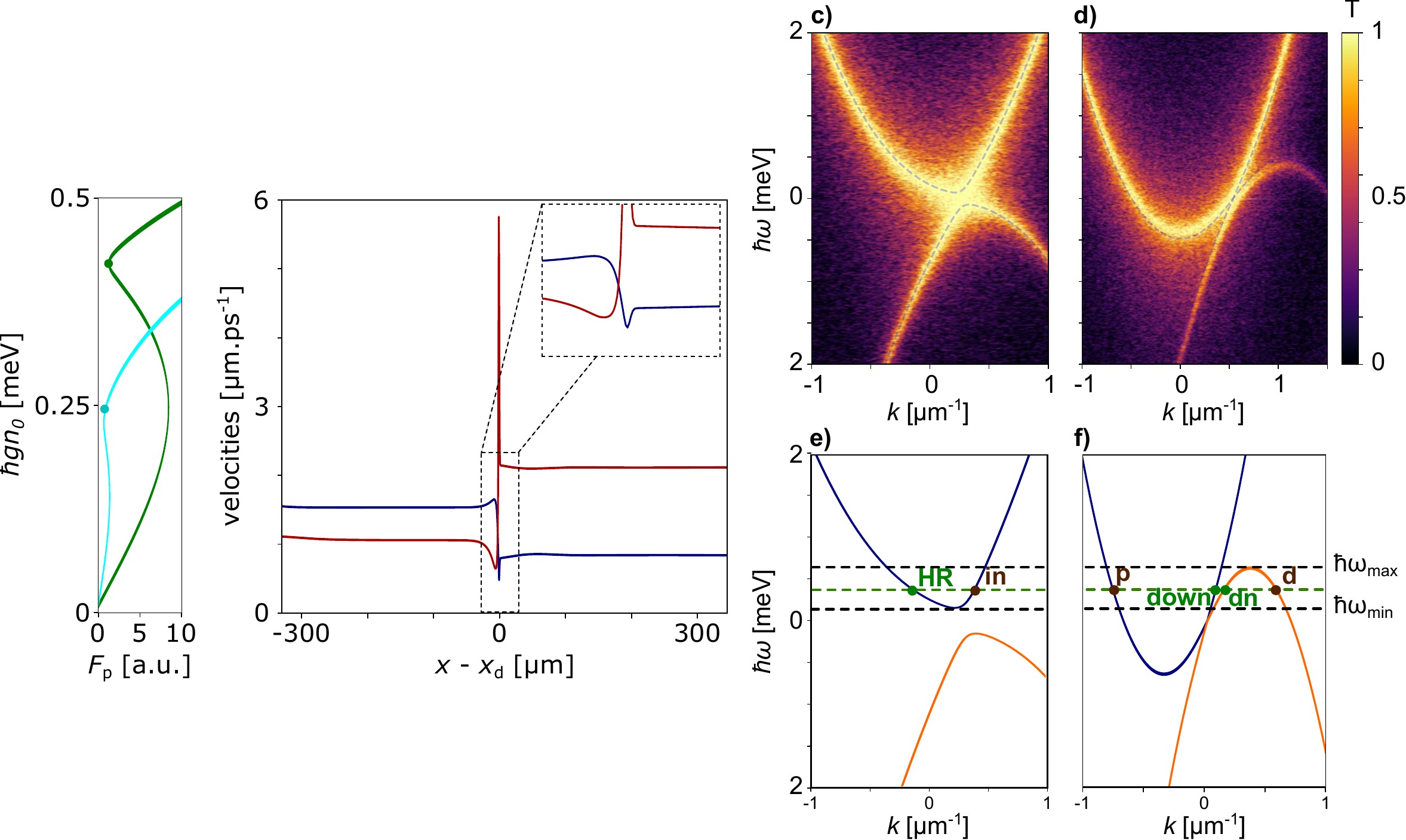}
    \caption{\textbf{Polariton mean-field}. Mean-field profile with downstream density support.
    \textbf{a)} Bistability loop for $k_\mathrm{up}=\SI{0.27}{\per\micro\meter}$ (green) and $k_\mathrm{down}=\SI{0.539}{\per\micro\meter}$ (turquoise).
    \textbf{b)} Bogoliubov sound speed $c_\mathrm{B}$ (blue) and background fluid velocity $v_0$ (red).
    \textbf{Dispersion of Bogoliubov excitations}. \textbf{c)}–\textbf{d)} The colour map shows the numerical dispersion relation, while the dashed lines show the LDA prediction~\eqref{eq:lfdisp}. \textbf{c)} upstream; \textbf{d)} downstream.
    \textbf{e)}, \textbf{f)}, LDA dispersion.
    Blue: positive-norm modes $\omega^{\mathrm B}_+$; orange: negative-norm modes $\omega^{\mathrm B}_-$.
    Brown dots: \textit{in} modes \textbf{in} (upstream), \textbf{p} and \textbf{d} (downstream); green dots: \textit{out} modes \textbf{HR} (upstream), \textbf{down} and \textbf{dn} (downstream).
    We stimulate the Hawking effect by injecting a finite-amplitude continuous-wave (CW) probe into the input mode \textbf{in}.}
    \label{fig:fig1}
\end{figure}

\section{The polaritonic simulator}
\label{sec:mean-field-linearisation}

Exciton–polaritons (polaritons) arise in planar semiconductor microcavities from the strong coupling between quantum-well excitons and cavity photons. Their hybrid light–matter character yields a small effective mass and sizeable nonlinearity, enabling collective phenomena akin to those in atomic Bose–Einstein condensates, but within optical platforms. Under coherent driving and dissipation, the dynamics of the polariton field $\psi(x, t)$ is governed by a driven–dissipative Gross–Pitaevskii equation (GPE)~\cite{carusotto_quantum_2013}
\begin{equation}
\label{eq:ddgpe}
i\hbar\,\partial_t \psi
=\Big[-\frac{\hbar^2}{2m^\ast}\partial_x^2+V(x)+\hbar g\,|\psi|^2-\frac{i\hbar\gamma}{2}\Big]\psi
+\hbar F(x)\,e^{-i\omega_\tp t},
\end{equation}
with effective mass $m^\ast$, interaction constant $g>0$, loss rate $\gamma$, and a structured pump envelope $F(x)$ of frequency $\omega_\tp$.

For a pump of constant amplitude $F(x)\equiv F_\tp$ and a plane-wave drive of wave-vector $k_\tp$ (stationary in the rotating frame), the steady flow $\psi_0=\sqrt{\nbg}\,e^{i(k_\tp x-\omega_\tp t)}$ yields the equation of state
\begin{equation}
\label{eq:EOS}
\nbg\Big[(g\nbg-\deff(\vbg))^2+\frac{\gamma^2}{4}\Big]=\frac{|F_\tp|^2}{\hbar^2}.
\end{equation}
Here $\deff(\vbg)=\omega_\tp-\omega_0-\frac{m^\ast \vbg^2}{2\hbar}$ is the effective detuning, and the hydrodynamic velocity is $\vbg=\hbar\,\partial_x\phi/m^\ast=\hbar k_\tp/m^\star$.
When $\deff/\gamma>\sqrt{3}/2$, the input–output curve is bistable (S-shaped) with a stable upper branch, used below as the high-density state.

We study collective excitations by linearising around a stationary background with the ansatz
\begin{equation}
\label{eq:lin-ansatz}
\psi(x,t)=e^{i(k_\tp x-\omega_\tp t)}\Big[\sqrt{\nbg(x)}+\delta\psi(x,t)\,e^{-\gamma t/2}\Big],\qquad
\delta\psi=u(x)\,e^{-i\omega t}+v^\ast(x)\,e^{+i\omega t}.
\end{equation}
In a spatially inhomogeneous system (spatially varying $\vbg(x)$ and $\cs(x)$), the Bogoliubov spinor $\begin{pmatrix}u\\ v\end{pmatrix}$ obeys~\cite{falque_polariton_2025}
\begin{equation}
\label{eq:BdG}
\Lbdg
\begin{pmatrix}u\\ v\end{pmatrix}
=\omega
\begin{pmatrix}u\\ v\end{pmatrix},\quad
\Lbdg=
\begin{pmatrix}
-\hbar\deff(\vbg)+2\hbar g \nbg + D & \hbar g \nbg\\[2pt]
-\hbar g \nbg & -\big[-\hbar\deff(\vbg)+2\hbar g \nbg + D\big]^{\!*}
\end{pmatrix},
\end{equation}
with the differential operator $D=-\frac{\hbar^2}{2m^\ast}\partial_x^2\;-\;i\hbar\,\vbg(x)\,\partial_x\;-\;i\frac{\hbar}{2}\,\partial_x \vbg(x)$, where the last term is local and purely imaginary; it vanishes in homogeneous regions and is appreciable only where $\vbg(x)$ varies significantly (e.g. near the horizon)~\cite{guerrero_multiply_2025}.
In what follows, we employ a local-density approximation (LDA) to obtain analytic laboratory-frame dispersions; within the LDA we neglect $\partial_x v_0$ and treat $n_0,v_0$ as piecewise constant when deriving \eqref{eq:lfdisp}.

We identify a \emph{Killing horizon} at the position where $|v_0|=\cs$.
In a slowly varying inhomogeneous flow, applying the LDA to \eqref{eq:BdG} yields the laboratory-frame dispersion relation
\begin{equation}
\label{eq:lfdisp}
\omega_\pm^{\mathrm B}(k;x)=\vbg(x)\,k\;\pm\;\sqrt{\Big[\frac{\hbar k^2}{2m^\ast}\Big]^2+\cs^2(x)\,k^2+\cs^2(x)\,\frac{m_{\mathrm{det}}^2(x)}{\hbar^2}},
\end{equation}
where the Bogoliubov light-cone velocity $\cs$ sets the causal bound $|\partial\omega/\partial k|\le\cs(x)$ and the mass parameter $m_{\mathrm{det}}$ encodes the local mass gap~\cite{falque_polariton_2025}:
\begin{equation}
\label{eq:cb-mdet}
\cs(x)=\sqrt{\frac{\hbar\,[2g\nbg(x)-\deff(\vbg(x))]}{m^\ast}},\qquad
m_{\mathrm{det}}(x)=m^\ast\,\frac{\sqrt{[g\nbg-\deff][3g\nbg-\deff]}}{\,2g\nbg-\deff\,}\Bigg|_{x}.
\end{equation}
The effective geometry is captured by the Painlevé--Gullstrand line element~\cite{visser_acoustic_1998}
\begin{equation}
\label{eq:PG}
ds^2=\big[\cs^2(x)-\vbg^2(x)\big]\,dt^2-2\,\vbg(x)\,dx\,dt-dx^2.
\end{equation}
As in all analogue simulators, the presence of a horizon is signalled by the excitation of negative-norm modes $\omega^{\mathrm B}_-$ at positive frequencies, i.e. negative-energy waves~\cite{falque_polariton_2025}.
\textit{Notation.} We denote the outside positive-norm channel by $u_{+}$ and the inside channels by $d_{+}$ (positive) and $d_{-}$ (negative).

\section{Numerical simulation of the mean-field and Bogoliubov spectrum}\label{sec:system}
We now simulate polariton dynamics in a one-dimensional (1D) sample (e.g. a wire or effective 1D propagation).
We use \textit{Julia} to integrate the (1+1)D GPE with a split-step scheme on a 2048-point grid (cavity length $\SI{800}{\micro\meter}$), with absorbing boundary conditions.
As in~\cite{jacquet_analogue_2022,jacquet_quantum_2023}, we take $m^\ast=5\times 10^{-35}\SI{}{\kilo\gram}$, $\hbar\omega_0=\SI{1473.36}{\milli\electronvolt}$, $\hbar\gamma=\SI{47}{\micro\electronvolt}$ and $\hbar g=3\times 10^{-4}\SI{}{\milli\electronvolt\micro\meter}$.
$V(x)$ is an attractive Gaussian potential (height $-\SI{0.85}{\milli\electronvolt}$, width $\SI{0.75}{\micro\meter}$) centred at $x_\mathrm{d}=\SI{400}{\micro\meter}$.
We refer to it as ``the defect''.

We employ a structured pump: high constant intensity for $x<x_\mathrm{d}-\SI{7}{\micro\meter}$ and low constant intensity for $x>x_\mathrm{d}-\SI{7}{\micro\meter}$.
The pump frequency is $\hbar\omega_\tp=\SI{1473.85}{\milli\electronvolt}$, while its wave-vector is structured as $k_\mathrm{up}=k_\tp(x<x_\mathrm{d}-\SI{7}{\micro\meter})=\SI{0.27}{\per\micro\meter}$ and $k_\mathrm{down}=k_\tp(x>x_\mathrm{d}-\SI{7}{\micro\meter})=\SI{0.539}{\per\micro\meter}$ (this specific value of $k_\mathrm{down}$ was chosen to match the phase across the horizon and to produce an approximately flat downstream density; see the appendix of~\cite{jacquet_analogue_2022} for details).

Fig.~\ref{fig:fig1} \textbf{a)} shows optical bistability for effective detunings $\hbar\delta(v_\mathrm{up})=\SI{0.39}{\milli\electronvolt}$ (green) and $\hbar\delta(v_\mathrm{down})=\SI{0.12}{\milli\electronvolt}$ (cyan).
Upstream, the pump intensity $F_{\tp,\mathrm{up}}$ is set $0.08\%$ above the turning point (green dot) to stabilise a steady state.
Downstream, in the supported case, $F_\tp(x>x_\mathrm{d}-\SI{7}{\micro\meter})$ is $16\%$ above the turning point (cyan dot).
\emph{These percentages are not directly comparable:} $0.08\%$ is an input-intensity offset on the upstream S-curve, whereas $16\%$ corresponds to an effective-detuning offset in the downstream supported region; because the S-curve slopes differ by $\mathcal{O}(10^2)$ across the horizon, a small upstream fractional shift maps to a much larger downstream relative offset.

Fig.~\ref{fig:fig1} \textbf{b)} shows $\vbg$ (red) and $\cs$ (blue).
As the fluid flows from left to right with increasing velocity, by analogy with a waterfall we refer to the high-density region as upstream and the low-density region as downstream.
Downstream, $v_\mathrm{down}$ reaches a constant value of $\SI{2.07}{\micro\meter\per\pico\second}$ set by $k_\mathrm{down}$ and exceeds $c_\mathrm{B}$, which is supported at a constant value of $\SI{0.81}{\micro\meter\per\pico\second}$.
The attractive defect causes a dip in the sound speed (equivalently, in the density) at $x_\mathrm{d}$, accompanied by a spike in the fluid velocity under quasi-conservation of the current (see the inset in Fig.~\ref{fig:fig1} \textbf{b)}).
Section~\ref{sec:QNMtheo} discusses the impact of these narrow features on the emission spectrum.

We probe the Bogoliubov dispersion by adding low-amplitude white noise at each time step of the simulation and by performing space–time Fourier transforms to obtain the $(k,\omega)$ spectrum shown in Fig.~\ref{fig:fig1}.
We restrict the spatial Fourier transform to obtain \emph{local} spectra in the upstream (Fig.~\ref{fig:fig1}\,\textbf{c}) and downstream (Fig.~\ref{fig:fig1}\,\textbf{d}) regions.
As the integration time is arbitrary, we normalise the spectral intensity to its maximum, enabling qualitative comparison only.
Upstream (Fig.~\ref{fig:fig1}\,\textbf{c}), the spectrum is slightly gapped, as expected for a pump intensity near but not at the bistability turning point.
A comparable gap is visible downstream when the fluid density is supported by a finite pump (Fig.~\ref{fig:fig1}\,\textbf{d}).

Assuming piecewise-homogeneous density and flow in each spatial region, the LDA gives the real part of the spectrum~\eqref{eq:lfdisp}.
We see in Fig.~\ref{fig:fig1} \textbf{c)} that the LDA captures the spectrum well when the density is high.
When the density is low (Fig.~\ref{fig:fig1}\,\textbf{d}), the LDA slightly overestimates the Mach number, yet the agreement with numerical data remains good (in line with spectral measurements in inhomogeneous media in~\cite{falque_polariton_2025,guerrero_multiply_2025}).
In what follows, we base our discussion on the LDA spectrum in Fig.~\ref{fig:fig1}\,\textbf{e},\textbf{f}.

Because of the Doppler effect, the downstream transcritical flow shifts $\omega^{\mathrm B}_-$ solutions (orange) to positive frequencies up to $\omega_\mathrm{max}$.
Within $\omega_\mathrm{min}<\omega<\omega_\mathrm{max}$ (bounded below by the upstream gap $\omega_\mathrm{min}$), there are three \emph{input} modes (brown): \textbf{in} ($\omega^{\mathrm B}_+$) propagating towards the horizon upstream, and \textbf{p} ($\omega^{\mathrm B}_+$) and \textbf{d} ($\omega^{\mathrm B}_-$) propagating against the supercritical flow downstream.
There are also three \emph{output} modes (green): \textbf{HR} ($\omega^{\mathrm B}_+$) propagating away from the horizon upstream, and \textbf{down} ($\omega^{\mathrm B}_+$) and \textbf{dn} ($\omega^{\mathrm B}_-$) propagating inside the horizon downstream.
Thus, over $\omega_\mathrm{min}<\omega<\omega_\mathrm{max}$, positive- and negative-norm modes coexist at positive frequencies.
This signals the formation of a Killing horizon for the Bogoliubov modes~\cite{falque_polariton_2025}.

\section{The stimulated Hawking Effect}\label{sec:HEtheo}
The Hawking effect arises from the scattering of incoming waves at the horizon, which mixes $\omega^{\mathrm B}_+$ and $\omega^{\mathrm B}_-$ solutions of Eq.~\eqref{eq:lfdisp} via the scattering matrix $S$.
A key feature is its pseudo-unitarity, dictated by conservation of the Bogoliubov scalar product,
\begin{equation}\label{eq:scalarproduct}
\langle \phi_1 | \phi_2 \rangle = \int dx\, \left(u_1^* u_2 - v_1^* v_2\right),
\end{equation}
which is indefinite and classifies modes into positive- and negative-norm branches, $\omega^{\mathrm B}_+$ and $\omega^{\mathrm B}_-$, respectively~\cite{castin_lecture_notes}. 
We now outline how dispersion controls scattering at the horizon and, in turn, how Hawking-type amplification arises.

The dispersion in Fig.~\ref{fig:fig1}\,\textbf{e},\textbf{f} implies that the properties of $S$ depend on frequency:
\begin{itemize}
    \item $\omega<\omega_\mathrm{min}$: only downstream channels propagate; \textbf{p} and \textbf{d} are incoming, while \textbf{down} and \textbf{dn} are outgoing. Equation~\eqref{eq:scalarproduct} leads to anomalous mixing described by a $2\times2$ matrix obeying $S^\dagger \mathrm{diag}(1,-1) S=\mathrm{diag}(1,-1)$;
    \item $\omega_\mathrm{min}<\omega<\omega_\mathrm{max}$: there are three input modes (\textbf{in}, \textbf{p}, \textbf{d}) and three output modes (\textbf{HR}, \textbf{down}, \textbf{dn}). \textbf{HR} is the unique upstream outgoing channel (Hawking radiation). Pseudo-unitarity reads $S^\dagger \mathrm{diag}(1,1,-1) S=\mathrm{diag}(1,1,-1)$;
    \item $\omega>\omega_\mathrm{max}$: only positive-norm channels remain, and scattering reduces to a two-port unitary process.
\end{itemize}

For $\omega<\omega_\mathrm{max}$, pseudo-unitarity entails mixing of creation and annihilation operators.
For example, when $\omega_\mathrm{min}<\omega<\omega_\mathrm{max}$,
\begin{equation}\label{eq:smatrix}
\begin{pmatrix}
\hat{a}_{\text{\textbf{HR}}} \\ \hat{a}_{\text{\textbf{down}}} \\ \hat{a}_{\text{\textbf{dn}}}^\dagger
\end{pmatrix}
= S \begin{pmatrix} \hat{a}_{\text{\textbf{in}}} \\ \hat{b}_{\text{vac,\textbf{p}}} \\ \hat{b}_{\text{vac,\textbf{d}}}^\dagger \end{pmatrix}.
\end{equation}
In the language of Gaussian quantum optics, this mixing is a concatenation of two symplectic operations: (i) a two-mode squeezer, responsible for pair creation across the horizon; (ii) a beam splitter, describing partial transmission through the effective potential barrier.

The probe cannot be injected from upstream for $\omega<\omega_\mathrm{min}$.
For $\omega>\omega_\mathrm{min}$, a coherent monochromatic probe $\ket{\eta_{\text{\textbf{in}}}}$ can be injected from upstream.
Considering that all other positive- and negative-norm input modes are in the vacuum state, the input state is $\ket{\rho_{\text{in}}}=\ket{\eta_{\text{\textbf{in}}}}\otimes\ket{0_{\mathrm{\textbf{p}}}}\otimes\ket{0_{\mathrm{\textbf{d}}}}$.
For $\omega_\mathrm{min}<\omega<\omega_\mathrm{max}$, Eq.~\eqref{eq:smatrix} shows that the incoming excitation is scattered into the reflected upstream mode (\textbf{HR}) and into two transmitted downstream modes (\textbf{down}, \textbf{dn}).
Crucially, conservation of the Bogoliubov scalar product implies that population of \textbf{dn} is necessarily accompanied by a net increase of the energy flux in co-propagating positive-norm channels, i.e. amplification.
For $\omega>\omega_\mathrm{max}$, unitarity of the $2\times2$ $S$-matrix forbids amplification, and the interface acts as a simple beam splitter.

Although the simulations use classical waves, the gain originates from mixing between positive- and negative-norm channels—the classical counterpart of parametric (squeezing) processes.
In particular, the gain is frequency-dependent and, in our configuration, peaks near the near-horizon resonance frequency $\Omega_{\mathrm{qnm}}$ (see Fig.~\ref{fig:fig2}\textbf{e}).
However, like other superradiant processes~\cite{delhom_entanglement_2024}, although $\omega$ is conserved, $k$ is not.

\section{Quasinormal modes and resonant transmission}\label{sec:QNMtheo}

Quasinormal modes (QNM) are the natural resonances of open-wave systems: they solve linearised equations with purely outgoing boundary conditions and have complex frequencies
\(\tilde\omega_{\mathrm{qnm}}=\Omega_{\mathrm{qnm}}-i\,\Gamma_{\mathrm{qnm}}/2\)~\cite{berti_quasinormal_2004,lalanne_light_2018}.
They enter our problem as poles of the horizon scattering matrix and control the long-lived response of the driven–dissipative polariton fluid.

In the present configuration, the attractive defect generates a narrow density dip and a companion spike in the flow velocity just inside the horizon (Fig.~\ref{fig:fig1}\,\textbf{b}, inset).
This forms a short, leaky resonator bounded by higher-density “shoulders”: impedance mismatches on either side partially reflect Bogoliubov waves while allowing leakage into the upstream and downstream asymptotic regions.
The QNMs are the resonant standing-wave solutions located in this region and radiating outward~\cite{jacquet_quantum_2023}: here, the QNMs are zero-norm modes composed of a localized negative-norm mode coupled to propagating positive-norm modes on either side of the horizon.
Semiclassically, they satisfy a lossy round-trip condition, which fixes a complex eigenfrequency; equivalently, they appear as simple poles of the relevant scattering amplitudes.
In practice, the real part sits just above the upper edge of the horizon frequency interval, where the downstream negative-norm channel closes. The imaginary part measures radiative leakage and is here comparable to intrinsic polariton losses~\cite{jacquet_quantum_2023}.

Consider a weak monochromatic probe injected in the upstream input mode \(\mathbf{in}\).
For \(\omega>\omega_\mathrm{max}\) the interface reduces to a two-port unitary scatterer between the incoming set \(\{\mathbf{in},\mathbf{p}\}\) and the outgoing set \(\{\mathbf{HR},\mathbf{down}\}\) (no negative-norm channel is available), so there is no net amplification, only redistribution between reflection and transmission.
In the single-sided drive used here, only \(\mathbf{in}\) is populated; the internal incidence channel \(\mathbf{p}\) is left in vacuum.
In this regime, the complex transmission coefficient in \(\mathbf{down}\) acquires a Breit--Wigner contribution~\cite{lalanne_light_2018}, $
t_{\mathrm{in}\to \mathrm{down}}(\omega)\;\simeq\;t_{\mathrm{bg}}(\omega)\;+\;\frac{\alpha}{\omega-\Omega_{\mathrm{qnm}}+i\,\Gamma_{\mathrm{qnm}}/2}$,
with a \(\pi\)-phase slip across \(\Omega_{\mathrm{qnm}}\) and a peak height/width governed by \(Q=\Omega_{\mathrm{qnm}}/\Gamma_{\mathrm{qnm}}\).
The corresponding intensity exhibits a narrow maximum at $\omega=\Omega_{\mathrm{qnm}}$, i.e. a frequency-resolved spectroscopy line outside the spontaneous-emission band.
This mechanism differs from ordinary tunnelling: It is mediated by a long-lived intermediate state set by the near-horizon geometry.

In momentum-resolved measurements, this appears as a sharp, frequency-selected enhancement along the upstream and downstream $\omega^{\mathrm B}_{+}$ loci, without concomitant growth of \textbf{dn} at the same \(\omega\) (the latter being kinematically closed)~\footnote{For completeness, we summarise the frequency-dependent availability of asymptotic channels and the role of the local Doppler shift of the $\omega^{\mathrm B}_-$ branch near the horizon.
For $\omega<\omega_\mathrm{min}$ only downstream channels propagate; for $\omega_\mathrm{min}<\omega<\omega_\mathrm{max}$, mixed-norm scattering is possible with one upstream outgoing channel (HR) and two downstream outgoing channels (down, dn); for $\omega>\omega_\mathrm{max}$ only positive-norm channels persist asymptotically.
Locally, the Doppler spike can bring parts of $\omega^{\mathrm B}_-$ to positive laboratory frequency, enabling norm-mixing inside the near-horizon resonator; asymptotically, however, \textbf{dn} closes above $\omega_\mathrm{max}$, and scattering reduces to a two-port unitary process.}.
This separation in $(k,\omega)$ mirrors the outgoing-channel mapping discussed earlier and provides a practical diagnostic for experiments.

The QNM therefore plays a dual role: Sets a privileged frequency just above \(\omega_\mathrm{max}\) at which probe transmission and reflection are resonantly enhanced and rotate in phase, and provides a geometric handle on the near-horizon structure (through \(\Omega_{\mathrm{qnm}}\) and \(\Gamma_{\mathrm{qnm}}\)).
We now extract \(\tilde\omega_{\mathrm{qnm}}\) from the linear problem and verify the predicted peak and phase behaviour of \(t_{\mathrm{in}\to \mathrm{down}}(\omega)\) numerically.

\section{Numerical simulation of scattering at the horizon}\label{sec:scatHE}
We now discuss the scattering of a low-amplitude continuous-wave (CW) probe sent towards the horizon from upstream, i.e. in mode \textbf{in}.
The probe amplitude is set to $|F_\tpr|=0.1\%\,|F_{\tp, \mathrm{up}}|$ so as to minimally perturb the mean field.
The probe is a $\SI{12}{\micro\meter}$ Gaussian centred on $x_\tpr=x_\mathrm{d}-\SI{100}{\micro\meter}$.
Bogoliubov excitations created by the probe propagate toward the horizon with group velocity $v_g(k_\tpr)=\frac{\partial\omega}{\partial k}$.

The simulation proceeds as follows: we first integrate the GPE with the source term $F_{\tp, \mathrm{up}}$ only, until a steady state is reached.
For each $k_\tpr,\omega_\tpr$, we then add a second source term $F_\tpr$ with $\omega_\tpr=\omega(k_\tpr)$, and allow the system to relax to a new steady state.
We then continue the time evolution and perform space--time Fourier transforms of $\psi-\psi_0$ to obtain amplitudes $A(k,\omega)$ in the up- and downstream regions, retrieved by windowed spatial Fourier analysis (Hann window).

This procedure effectively simulates the scattering of \textbf{in} amplitude into the outgoing modes \textbf{HR}, \textbf{dn} and \textbf{down}, as in an experiment.
Because frequency resolution is obtained by time-domain Fourier analysis, the $S$-matrix coefficients (interface transmission and reflection) cannot be extracted directly from these data.

\begin{figure}[ht]
    \centering
    \includegraphics[width=\linewidth]{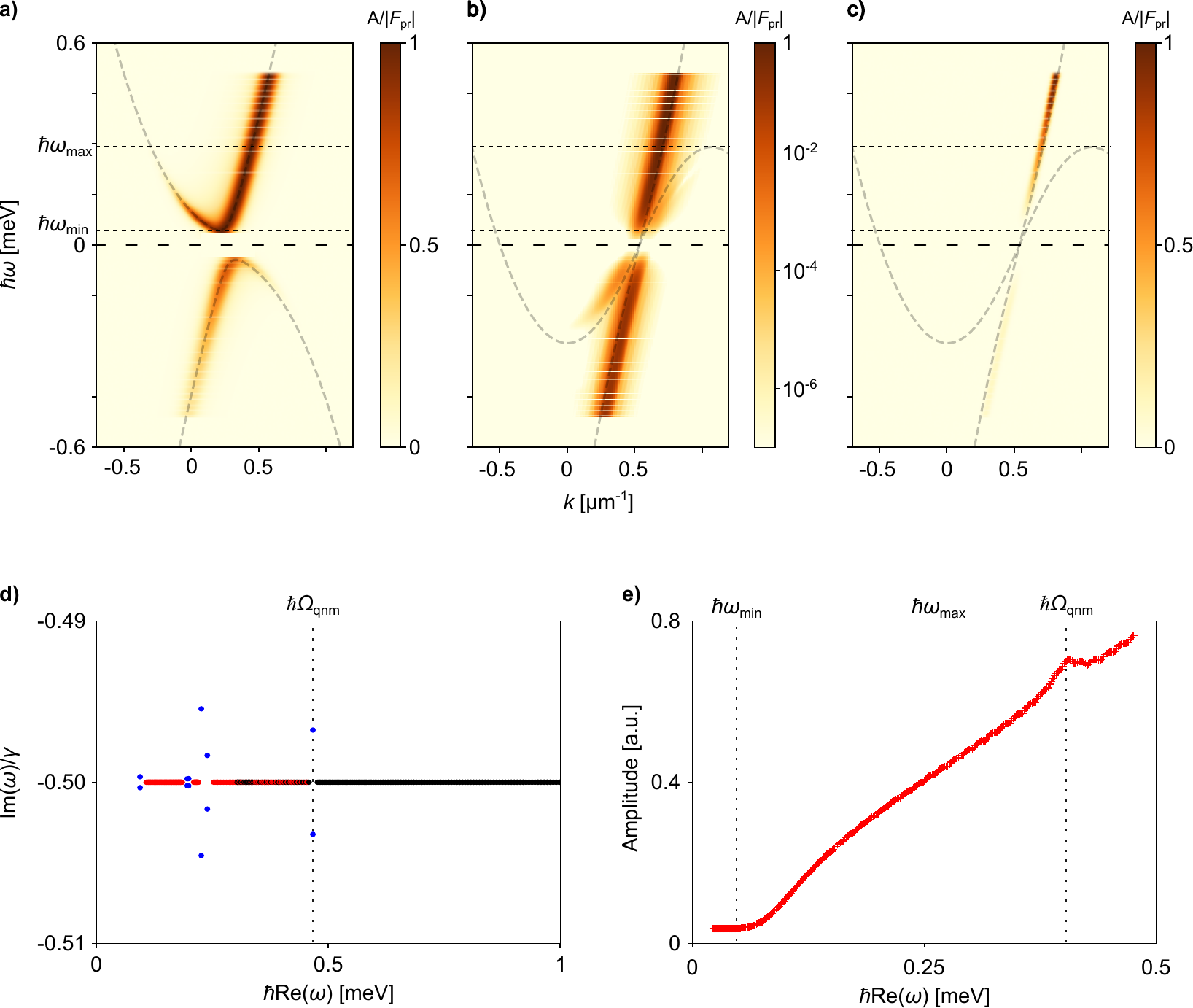}
    \caption{\textbf{Numerical simulation of scattering}.
    Amplitude is injected in mode \textbf{in} at $k_\tpr,\,\omega_\tpr$ and allowed to scatter at the horizon.
    Amplitudes are normalised to the probe amplitude.
    All $\omega_\tpr$ slices are collated to form the spectra in each region.
    \textbf{a)} amplitude upstream (linear scale); \textbf{b)} amplitude downstream (log scale); \textbf{c)} amplitude downstream (linear scale).
    Dashed lines: LDA dispersion~\eqref{eq:lfdisp}.
    \textbf{Spectral properties of the quasinormal mode}.
    \textbf{d)} Spectrum of Bogoliubov excitations. Red: negative-norm; black: positive-norm; blue: zero-norm. Zero-norm modes at low frequency are spurious artefacts of the numerical boundaries and are localised outside the near-horizon region. The response near $\omega_\mathrm{qnm}$ is a \emph{damped QNM} response of the Bogoliubov field.
    \textbf{e)} Amplitude in mode \textbf{down}.}
    \label{fig:fig2}
\end{figure}

Figure~\ref{fig:fig2} shows the simulation results.
The amplitude is normalised to the input amplitude.
Gaps in the spectra reflect cases where the resonance search fails at a given $k_\tpr,\omega_\tpr$, in which case the run is omitted.
Dashed lines show the LDA dispersion.
Reflected amplitudes broadly follow the LDA prediction, whereas downstream amplitudes are modulated by a Fano-like interference arising from the interplay between finite-windowing and the spectral proximity of downstream branches.
This modulation slightly shifts the amplitude profiles along $k$.
A similar effect may occur in experimental data; however, the overall shape of the transmission spectrum is unaffected and remains a robust observable of the underlying physics.

Consider the top row of Fig.~\ref{fig:fig2}: \textbf{a)} shows the upstream amplitude (linear scale), while \textbf{b)} (\textbf{c)}) shows the downstream amplitude on logarithmic (linear) scale.
In \textbf{a)}, the probe is injected from $\omega_\mathrm{min}$ (below which injection is not possible).
Comparing \textbf{a)} and \textbf{c)}, the amplitude is predominantly reflected into \textbf{HR} up to $\hbar\omega\approx \SI{0.25}{\milli\electronvolt}$; at higher frequencies, it is chiefly transmitted into \textbf{down}.
In \textbf{b)}, a finite amplitude is also transmitted into \textbf{dn}, the hallmark of Hawking-type amplification.

In Fig.~\ref{fig:fig2}\,\textbf{b}, non-negligible amplitude appears in \textbf{down}$^\star$ and \textbf{dn}$^\star$ (at negative $\omega$), as expected in Bogoliubov theory.
Because \textbf{dn} is a negative-norm mode with $|v|>|u|$, the field weight resides mainly in the conjugate component; consequently the \(\mathbf{dn}^\star\) trace is generically stronger than \textbf{dn} itself, a trend enhanced when downstream support separates the $\omega_\pm$ branches.

For $\omega>\omega_\mathrm{max}$, reflection drops to zero and the probe amplitude is transmitted downstream into \textbf{down}.
As discussed in Section~\ref{sec:HEtheo}, at these frequencies the interface between sub- and supercritical flows is no longer a horizon (no negative-norm mode is available at positive frequencies) and acts as a simple beam splitter.
Scattering is then dominated by transmission in \textbf{down}, with a very small reflected amplitude (of order $10^{-3}$ here), and no amplification occurs.

\section{Numerical observation of resonant scattering}\label{sec:scatqnm}
Section~\ref{sec:HEtheo} established that for $\omega>\omega_\mathrm{max}$ scattering at the interface is unitary, and Section~\ref{sec:scatHE} showed that the input amplitude is then transmitted downstream into \textbf{down}.
Section~\ref{sec:QNMtheo} further anticipates a QNM signature: an enhancement of transmission into \textbf{down}.

The spectral properties of the QNM are shown in Fig.~\ref{fig:fig2}.
In \textbf{d)}, we diagonalise the Bogoliubov matrix and obtain its eigenmodes.
Positive-norm modes are shown in black, negative-norm modes in red, and zero-norm modes in blue.
Zero-norm modes at low frequency ($<\SI{0.2}{\milli\electronvolt}$) are artefacts of the periodic boundary conditions in the diagonalisation.
The QNM is the only zero-norm mode at higher frequencies, just above $\omega_\mathrm{max}$.
Here, $\Gamma_\mathrm{qnm}/2$ is only slightly different from the intrinsic polariton losses.
Thus, linear losses set the QNM linewidth ($\Gamma_{\mathrm{qnm}}$) and prevent dynamical runaway inside the near-horizon resonator.
In \textbf{b)}, we show the amplitude in \textbf{down}, retrieved from Fig.~\ref{fig:fig2}\,\textbf{e}.
The amplitude is zero before $\omega_\mathrm{min}$ and then increases monotonically with $\omega$, with a peak at $\Omega_\mathrm{qnm}$.
This confirms resonant transmission.

Repeating the calculations without downstream density support yields the same qualitative and quantitative features.
Beyond the QNM line itself, all other observables (mode splitting, channel mapping, frequency windows) are likewise robust to the presence or absence of support.
This robustness indicates that the QNM is not an artefact of boundary pumping, but a structural feature of the effective spacetime geometry.

In summary, the spectral peak at $\omega=\Omega_\mathrm{qnm}$ confirms that the QNM arises from the internal geometry of the polariton fluid rather than from extrinsic gain.

\section{Conclusions}
We have theoretically validated the feasibility of observing stimulated Hawking radiation in a realistic polariton platform using frequency-resolved probe spectroscopy.
Horizon scattering in our driven polariton fluid shows high reflectivity into \textbf{HR} at low frequencies, accompanied by transmission into \textbf{dn} (the negative-norm partner), while the downstream amplitude is dominated by the positive-norm witness mode \textbf{down}~\cite{isoard_departing_2020,porrotunneling2024}.
With downstream density support, the separation of $\omega^\pm$ branches by the mass gap, and the concomitant increase in interaction energy, enhance the visibility of \textbf{dn}.
This support is experimentally feasible, strengthening the case for polariton platforms as analogue quantum-field simulators on tailored curved spacetimes and enabling systematic studies of how horizon configurations imprint the Hawking spectrum~\cite{parentani_entanglementhorizon_2010,Barcelo:2010xk,finazzi_robustness_2011,finazzi_spectral_2011,fabbri_rampup_2021,jacquet_quantum_2023,agullo_entanglement_2024,porrotunneling2024}.

In the present transcritical geometry, a narrow near-horizon resonator supports a quasinormal mode of the Bogoliubov field~\cite{jacquet_quantum_2023}.
Our simulations show a pronounced frequency-resolved peak in the stimulated transmission in \textbf{down} at $\omega=\Omega_{\mathrm{qnm}}$, outside the Hawking-emission interval.
This enables diagnosis of the effective geometry by transmission spectroscopy, an analogue of black-hole spectroscopy.
The QNM thus plays a dual role: it is both a dynamical signature of the Killing horizon and an active mediator of amplification.
Its frequency sets a privileged scale at which the near-horizon geometry becomes dynamically relevant, providing a resonant fingerprint of the engineered curved spacetime.
As the required conditions have already been demonstrated~\cite{falque_polariton_2025}, these effects should be accessible to current experiments, offering a platform for addressing open questions in near-horizon physics~\cite{york_dynamical_1983,hod_bohrs_1998,maggiore_qnm_2008,jacquet_quantum_2023}.

\section{Acknowledgements}
We thank Quentin Valnais, Malo Joly, and Tangui Aladjidi for assistance with the numerical code, and Marcos Gil de Olivera, Killian Guerrero, and Elisabeth Giacobino for insightful discussions on the dynamics of quantum fluids of light.
Our investigations were inspired in large part by Renaud Parentani's work on quantum fluids in analogue gravity; the references cited here reflect only a subset of his and his collaborators' contributions to the field.
M.J.J. and A.B. acknowledge funding from EU Pathfinder 101115575 Q-One, the DIM SIRTEQ project FOLIAGE, and CNRS via an 80$^\mathrm{prime}$ PhD studentship. A.B. acknowledges support from the Institut Universitaire de France.
IC  acknowledges financial support from Provincia Autonoma di Trento (PAT)  and from the National Quantum Science and Technology Institute through  the PNRR MUR project under Grant PE0000023-NQSTI, co-funded by the European Union – NextGeneration EU.

%

\end{document}